**Giant Transport Anisotropy in ReS$_2$ Revealed via Nanoscale Conducting Path Control**

Dawei Li,[1] Shuo Sun,[1] Zhiyong Xiao,[1] Jingfeng Song,[1] Ding-Fu Shao,[1] Evgeny Y. Tsymbal,[1] Stephen Ducharme,[1] Xia Hong[1*]

[1] Department of Physics and Astronomy & Nebraska Center for Materials and Nanoscience, University of Nebraska-Lincoln, Lincoln, Nebraska 68588-0299, USA

[*] Correspondence to: xia.hong@unl.edu

**Abstract**

The low in-plane symmetry in layered 1T'-ReS$_2$ results in strong band anisotropy, while its manifestation in the electronic properties is challenging to resolve due to the lack of effective approaches for controlling the local current path. In this work, we reveal the giant transport anisotropy in monolayer to four-layer ReS$_2$ by creating directional conducting paths via nanoscale ferroelectric control. By reversing the polarization of a ferroelectric polymer top layer, we induce conductivity switching ratio of $>1.5 \times 10^8$ in the ReS$_2$ channel at 300 K. Characterizing the domain-defined conducting nanowires in an insulating background shows that the conductivity ratio between the directions along and perpendicular to the Re-chain can exceed $5.5 \times 10^4$ in monolayer ReS$_2$. Theoretical modeling points to the band origin of the transport anomaly, and further reveals the emergence of a flat band in few-layer ReS$_2$. Our work paves the path for implementing the highly anisotropic 2D materials for designing novel collective phenomena and electron lensing applications.

Layered two-dimensional (2D) semiconductors such as black phosphorus and 1T'-rhenium disulfide (ReS$_2$) possess low in-plane symmetry, which leads to a rich spectrum of intriguing electronic and optical phenomena [1], including intrinsic band anisotropy [2], strongly anisotropic bound excitons and nonlinear optical responses [3-6], tunable hyperbolic plasmonics [7-9], large optical birefringence [10, 11], moiré superlattices [12-14], and multiferroic behaviors [15, 16]. The transition metal dichalcogenide (TMDC) ReS$_2$ is a direct band gap semiconductor, with the band gap ($E_g$) varying from 1.43 eV in monolayer samples to 1.35 eV in bulk [17]. It exhibits strong in-plane anisotropy between the directions along and perpendicular to the Re chains. Theoretical studies have shown that the band dispersions lead to highly direction-dependent mobility in ReS$_2$ [18, 19], while direct mapping of its angle-resolved transport remains challenging due to the lack of effective strategies to control the local current path [20, 21].

A promising approach to define reconfigurable, directional conduction paths in the 2D semiconductors is to leverage the nanoscale controllable polarization of an adjacent ferroelectric layer. In previous studies, ferroelectric domain patterning has been exploited to impose a wide range of functionalities in TMDCs [22, 23], including programmable homo- and hetero-junction states [24, 25] and photovoltaic effects [26], nanoscale excitonic modulation [27, 28], and nonlinear optical filtering [29]. Combining local polarization writing with the ferroelectric field effect enables the programming of nanoscale conduction paths within an insulating background, which can confine the local current flow in different pre-designed directions on the same sample. Unlike lithographically defined nanowires, the nonvolatile field effect approach is clean, reversible, and does not involve uncontrollable sample-to-sample variations.

In this work, we exploited nanoscale polarization control of a ferroelectric copolymer poly(vinylidene fluoride-trifluoroethylene) [P(VDF-TrFE)] top layer to probe the transport anisotropy in atomically thin ReS$_2$. By characterizing the domain-defined conducting nanowires in the insulating channel [Fig. 1(a)], we mapped out the angle-resolved conductance of single-layer (1L), bilayer (2L), and four-layer (4L) ReS$_2$ field effect transistors (FETs), which revealed a giant conductivity ratio of $5.5\times10^4$ between the directions along and perpendicular to the Re-chain [Fig. 1(b)]. The transport results can be well accounted for by the band anisotropy in conjunction with the electron-phonon scattering, as revealed by our first-principles density functional theory (DFT) modeling, which further points to the emergence of a flat band in the 4L ReS$_2$. Our study illustrates a powerful approach for resolving nanoscale electronic signatures of emergent band properties in van der Waals (vdW) materials, as well as presenting a promising material platform for realizing correlation-driven quantum phenomena and electron lensing applications.

We mechanically exfoliated monolayer and few-layer ReS$_2$ flakes from bulk single crystals onto the Gel-Films. The layer number was confirmed by atomic force microscopy (AFM, Bruker MultiMode 8) studies combined with the Raman frequency difference Δ between the modes I and III. The crystalline orientation of ReS$_2$ was identified by angle-resolved parallel-polarized Raman scattering measurements (Supplemental Materials) [30, 31]. Selected 1-4L ReS$_2$ flakes were transferred onto SiO$_2$ (290 nm)/doped Si substrates pre-patterned with Cr/Au (2 nm/10 nm) electrodes, forming FET devices. Next, we deposited 9 monolayers of P(VDF-TrFE) film on top of ReS$_2$ using the Langmuir-Blodgett (LB) technique [32] followed by a thermal annealing treatment at 135 ºC for 80 min (Supplemental Materials) [30]. Our previous studies have shown that the thermally treated P(VDF-TrFE) on ReS$_2$ forms close-packed single crystalline nanowires with the polar axis along the film normal [33]. To switch the ferroelectric polarization, a ±11 V DC bias ($V_{bias}$) was applied to a conductive PtIr-coated tip (Bruker SCM-PIC-V2), with domain writing controlled by the NanoMan program. The resulting domain structures were imaged using piezoresponse force microscopy (PFM). The electrical characterization was performed using the semiconductor parameter analyzer (Keysight B1500A) after domain patterning, while the sample was kept in the AFM during the entire process.

Figures 1(c)-(e) display the PFM phase images taken on a 1L ReS$_2$ FET, with the P(VDF-TrFE) top layer in the as-prepared, no poling state [Fig. 1(c)] and the uniformly patterned polarization down ($P_{\text{down}}$) [Fig. 1(d)] and up ($P_{\text{up}}$) [Fig. 1(e)] states, respectively. In all states, the sample exhibits linear source-drain current-voltage ($I_d$-$V_d$) relation, confirming the Ohmic characteristic (Supplemental Fig. S3) [30]. Figure 1(f) compares the $I_d$ versus back-gate voltage ($V_{\text{bg}}$) relation for these three states, showing the $P_{\text{down}}$ ($P_{\text{up}}$) state accumulates (depletes) electrons in ReS$_2$, as expected. The transfer curve in the $P_{\text{up}}$ state exhibits hole-doped characteristic, with $I_d$ decreasing with increasing $V_{\text{bg}}$. In contrast, the $P_{\text{down}}$ polarization of P(VDF-TrFE) introduces a high level of electron-doping in ReS$_2$, changing the dominant carrier type from *p*-type to *n*-type. In this state, the ReS$_2$ channel remains highly conductive over the entire $V_{\text{bg}}$–range, showing that the accumulated electron density well exceeds what can be effectively depleted by the SiO$_2$ back-gate. The current switching ratio between the $P_{\text{down}}$ and $P_{\text{up}}$ states reaches $5.3\times10^6$ at $V_{\text{bg}}$ = 7 V. A higher switching ratio may be expected at $+7\ V \leq V_{\text{bg}} \leq +40$ V, where the 1L ReS$_2$ becomes so insulating in the $P_{\text{up}}$ state that the current level is below the instrument resolution.

We also achieved nonvolatile current modulation in the 2L and 4L ReS$_2$ FET devices [Fig. 1(g) and Supplemental Fig. S3] [30]. In the 4L device, we extracted from the transfer curves a high current switching ratio of $1.5\times10^8$ at $V_{\text{bg}}$ = 12 V, which is among the highest values reported in ferroelectric-gated 2D FETs [34]. Compared with the polymorphous polymer films prepared via spin coating [21], where the net polarization is compromised due to the randomly oriented polar grains, the thermally treated LB films enable the remarkable level of doping modulation, which is critical for defining a highly insulating background. In the $P_{up}$ state, both 2L and 4L channels exhibit a transition from *p*-type to *n*-type transfer characteristics with increasing $V_{bg}$, indicating that the Fermi level is shifted close to the conduction band. Even though the fractional change of the doping level is expected to increase with decreasing channel thickness, we find the ferroelectric field effect is larger in the thicker samples. A likely reason is 1L ReS$_2$ is highly insulating in the $P_{up}$ state and cannot provide sufficient screening to P(VDF-TrFE). This results in a high depolarization field, which leads to incomplete polarization switching [23].

To create a directional conducting nanowire, we switched the polarization of P(VDF-TrFE) into the uniform $P_{up}$ state, setting the entire ReS$_2$ channel in a highly insulating state, and then wrote a line-shaped $P_{down}$ domain between the source and drain. Given the high current switching ratio, the nanowire conductance remains orders of magnitude higher than the insulating background over the entire $V_{bg}$-range. We worked with $P_{down}$ nanowires that are 300-400 nm wide. At this width range, electron conduction can be effectively confined to be along the nanowire direction, while the nanowire can sustain the 2D transport characteristic without a prominent edge contribution. Writing $P_{down}$ nanowire domains along different orientations thus enables angle-resolved conductance measurements on the same sample. Figure 2 shows the results obtained on a 1L ReS$_2$ FET using this approach. The *b*-axis of this sample is perpendicular to the channel orientation [Fig. 2(a)] [30]. We wrote a series of $P_{down}$ nanowires connecting the source and drain electrodes [Figs. 2(b)-2(e)]. Figure 2(f) shows the corresponding channel sheet conductance $\sigma = \frac{L}{W}\frac{I_d}{V_d}$, or 2D conductivity, as a function of $V_{bg}$ for the nanowires. Here *L* and *W* are the length and width of the nanowire, respectively, and the angle $\theta$ is defined with respect to the *b*-axis of ReS$_2$. It is evident that there are two distinct transfer characteristics for these four angles. For the nanowire along the directions of $\theta = 30°$ and $150°$, which are oriented close to the *b*-axis, the channel remains highly conductive with *n*-type characteristic over the entire $V_{bg}$-range. In contrast, for $\theta = 75°$ and $90°$, which are close to the direction perpendicular to the *b*-axis, the ReS$_2$ nanowires exhibit very low conductance, with the channel effectively turned into *p*-type at $V_{bg} < 20\,V$. At $V_{bg} = 20\,V$, ReS$_2$ exhibits up to $5.5 \times 10^4$-fold change in conductance between the directions of $\theta = 30°$ and $\theta = 75°$, which also corresponds to a change of carrier type from electron- to hole-doped behavior. An even larger transport anisotropy is expected between $V_{bg} = 20\,V$ and $35\,V$, where the channel is too insulating along $\theta = 75°$ direction to be measured.

To map out the angle-resolved transport close to the conduction band edge, we extracted the conductivity of nanowires in various directions at fixed $V_{bg}$. We chose a $V_{bg}$ at which the sample is at the threshold of being turned on and electron-doped in all directions. Figures 3(a) shows the polar plot of $\sigma$ for the 1L ReS$_2$ sample at $V_{bg} = 40\,V$. The sheet conductance measured at the angles $\theta = 30°$ (954 nS) is about 56 times of that at $\theta = 105°$ (17 nS). Similar electron conduction anisotropy has been observed in the 2L [Fig. 3(b)] and 4L [Fig. 3(c)] ReS$_2$ samples (Supplemental Materials) [30]. Despite the fluctuation of the data points, which can be affected by the local defects in the P(VDF-TrFE) top-layer close to the nanowire area, all samples exhibit high conductance in the vicinity of the *b*-axis ($\theta = 0°$) and low conductance when $\theta$ approaches $90°$. The anisotropy is strongly enhanced in the few-layer sample, with a giant anisotropic conductance ratio of about $1.7 \times 10^3$ observed in the 4L ReS$_2$ device between the directions of $\theta = 0°$ (496.5 nS) and $120°$ (0.3 nS) [Fig. 3(c)]. Unlike the modulation observed in Fig. 2(f), this change does not involve the change of carrier type, thus reflecting solely the electron transport anisotropy. Note that the observed angular dependence of $\sigma$ reflects a lack of mirror symmetry with respect to the $\theta = 0°$ axis. As 1T'-ReS$_2$ belongs to the space group $P\bar{1}$, it only has an inversion symmetry.

The conductance in ReS$_2$ is determined by the electron mobility $\mu = e\tau/m^*$, where $e$ is the electron charge, $\tau$ is the average scattering time induced by defects, impurities, or phonons, and $m^*$ is the effective mass determined by the band structure $E(\boldsymbol{k})$ as $m^* = \frac{\hbar^2}{\partial^2 E(\boldsymbol{k})/\partial \boldsymbol{k}^2}$. To understand the role of band dispersion on the anisotropic conduction, we performed the first-principles DFT calculations for the band structures of 1L [Fig. 4(a)], 2L [Fig. 4(b)], and 4L [Fig. 4(c)] ReS$_2$. The details of the calculation can be found in the Supplemental Materials [30, 35-38]. The result obtained for 1L ReS$_2$ is consistent with the previous report [20]. For all three layer numbers, the band dispersion along the $k_x$ direction (*b*-axis) at the conduction band minimum (CBM) exhibits a larger curvature compared with that along the $k_y$ direction [Figs. 4(d)-(f)], yielding a lighter $m^*$ along *b*-axis ($\theta = 0°$) and a heavier $m^*$ perpendicular to *b*-axis ($\theta = 90°$). The anisotropy of the band dispersion, on the other hand, is significantly enhanced with increasing layer thickness of ReS$_2$. For all layer thicknesses, the inversion symmetry of the space group $P\bar{1}$ has been preserved in the calculated $m^*$ (Supplemental Materials) [30].

Figures 4(g)-(i) show the normalized $1/m^*$ vs. $\theta$ relation superimposed onto the normalized conductance ($\sigma_{\text{norm}}$) data. For 1L ReS$_2$, $1/m^*$ shows a similar $\theta$ −dependence as the measured $\sigma$, *i.e.*, the maximum values appear at orientations close to $\theta = 0°$ and $180°$ and the minimum values appear close to $\theta = 90°$ and $270°$ directions [Fig. 4(g)]. The variation in $1/m^*$, however, cannot fully account for the relative change in $\sigma_{\text{norm}}$. Considering that the band anisotropy also affects the election-phonon scattering, we incorporated the contribution of phonon scattering in mobility $\mu$ by using the Takagi formula [39-41]:

$$\mu_i = \frac{e\hbar^3 C_i}{k_B T m_i^* m_d^* D_i^2}. \tag{1}$$

Here $i$ refers to the conduction direction, and $m_d^* = \sqrt{m_x^* m_y^*}$ is the density-of-state effective mass for an anisotropic electronic band. The deformation potential constant is defined as $D_i = \frac{\partial E_V}{\partial \varepsilon_i}$, where $E_V$ is the energy of the CBM and $\varepsilon_i$ is a strain applied along direction $i$. The 2D elastic modulus along the conduction direction is calculated using $C_i = \frac{2 \partial^2 E_{\text{total}}}{S_0 \partial \varepsilon_i^2}$, where $E_{\text{total}}$ is the total energy of ReS2 and $S_0$ is the area of the 2D ReS2 without strain. As shown in Fig. 4(g), the modeled $\mu$ vs. $\theta$ relation shows improved agreement with $\sigma_{\text{norm}}(\theta)$, indicating that the anisotropic conductance in 1L ReS2 is resulted from the convoluted effects of the anisotropic band dispersion and the phonon scattering. On the other hand, we find that $\sigma_{\text{norm}}(\theta)$ for the 2L [Fig. 4(h)] and 4L [Fig. 4(i)] ReS2 can be well explained by the calculated $1/m^*$ vs. $\theta$ relation. This is because the anisotropy of the conduction band dispersion is significantly enhanced in thicker ReS2 [Figs. 4(d)-(f)], so that the angular dependence of electron-phonon scattering plays a relatively minor role in the electron transport in 2L and 4L ReS2.

These findings are in sharp contrast with the previous studies of directional transport in ReS2 using a circular sample geometry with multiple point-contacts along various directions [20] or micron-sized conduction channels [21], where the experimentally extracted anisotropic conductance ratio is significantly smaller than the theoretical prediction. In both approaches, the local current path within the sample cannot be well controlled. The measurements thus collect current flows over a wide range of angle distributions, making it challenging to quantitatively analyze the directional conduction.

What is intriguing is the emergence of a nearly flat band in 4L ReS2 at the CBM along the $k_y$ direction [Fig. 4 (f)], leading to a drastic increase in $m^*$ for $\theta = 90°$ (Supplemental Materials), which can well account for the giant electron transport anisotropy observed in 4L ReS2 [30]. Such band dispersion highly resembles what is predicted for one-dimensional graphene superlattices (GSL) [42]. It is thus conceivable to design ReS$_2$-based novel electron lensing applications by either varying the layer numbers along the conducting path through assembling artificial ReS$_2$ heterostructures or integrating few-layer ReS$_2$ with other isotropic vdW materials. Compared with the GSL approach, which is challenging to realize due to difficulties in imposing nanoscale 1D periodic potential, few-layer ReS$_2$-based electron supercollimation builds upon the fabrication of vdW heterostructures and is free of lithography induced disorder and fluctuation. The flat band and the associated heavy effective mass also make few-layer ReS$_2$ a promising platform for hosting collective phenomena, such as magnetism and superconductivity.

In summary, we have resolved for the first time the giant transport anisotropy in mono- to few-layer ReS$_2$ by creating directional conducing paths through the nanoscale ferroelectric control, which reveals a 2D conductivity ratio exceeding $5.5\times10^4$ between the directions along and perpendicular to *b*-axis in 1L ReS$_2$. Our DFT calculations point to the band origin of this intriguing behavior, and reveal the emergence of a flat band in few-layer ReS$_2$. Our approach can be widely applied to other anisotropic vdW materials, providing a novel route for resolving nanoscale electronic signatures of emergent band properties and designing collective phenomena and electron lensing applications in vdW heterostructures.

**Acknowledgements**

We thank Xi Huang and Yongfeng Lu for the access to the Raman system. This work was primarily supported by the U.S. Department of Energy (DOE), Office of Science, Basic Energy Sciences (BES), under Award No. DE-SC0016153 (sample preparation and characterization, FET device fabrication and characterization). S.S. and S.D. acknowledge the support of the Nebraska

Center for Energy Sciences Research. The work of D.-F.S. and E.Y.T. was supported by the NSF Nebraska Materials Research Science and Engineering Center (MRSEC) under Grant No. DMR-1420645 (theoretical modeling). The research was performed in part in the Nebraska Nanoscale Facility: National Nanotechnology Coordinated Infrastructure and the Nebraska Center for Materials and Nanoscience, which are supported by the National Science Foundation under Award ECCS: 2025298, and the Nebraska Research Initiative.

# Figure 1

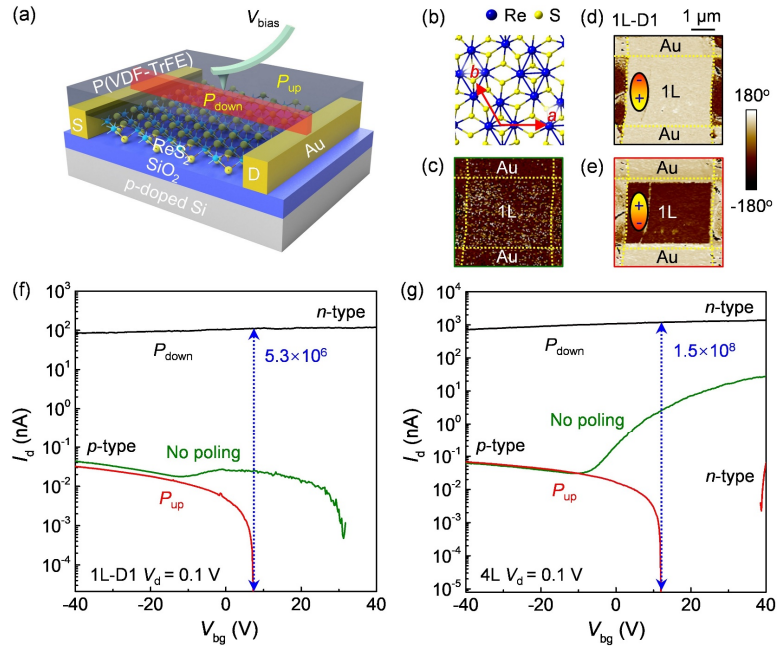

FIG. 1. (a) Device schematic. (b) Schematic crystal structure of $ReS_2$. (c-e) PFM phase images of a 1L $ReS_2$ FET (1L-D1) with the P(VDF-TrFE) top-gate (c) in the initial no poling state and uniformly polarized (d) $P_{down}$ and (e) $P_{up}$ states. (f-g) $I_d$ vs. $V_{bg}$ for (f) the same 1L $ReS_2$ FET and (g) a 4L $ReS_2$ in different polar states of P(VDF-TrFE).

Figure 2

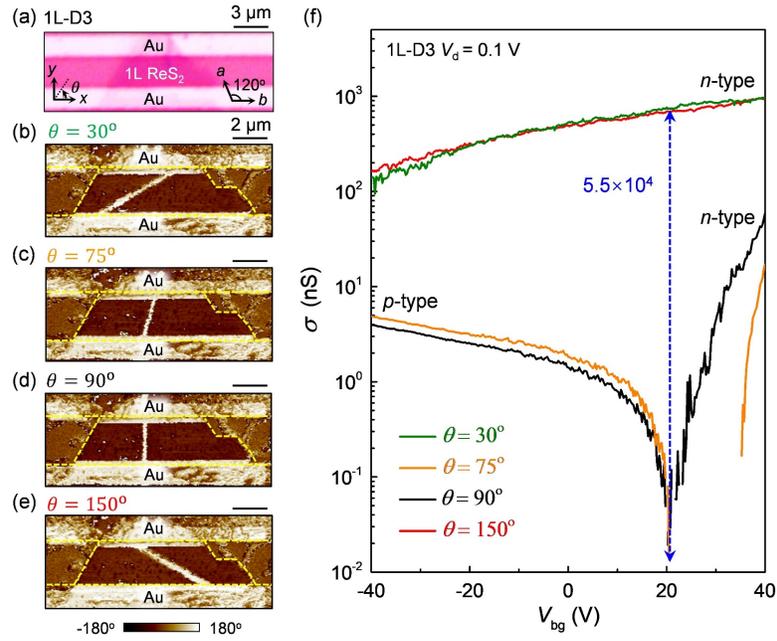

FIG. 2. (a) Optical image of a 1L ReS$_2$ device. Insets: laboratory coordinates (left) and crystalline axes of ReS$_2$ (right). (b-e) PFM phase images of $P_{down}$ nanowires created on the uniform $P_{up}$ background along (b) $\theta = 30°$, (c) $\theta = 75°$, (d) $\theta = 90°$, and (e) $\theta = 150°$ directions with respect to $b$-axis of ReS$_2$. (f) The corresponding $\sigma$ vs. $V_{bg}$.

Figure 3

Figure 3

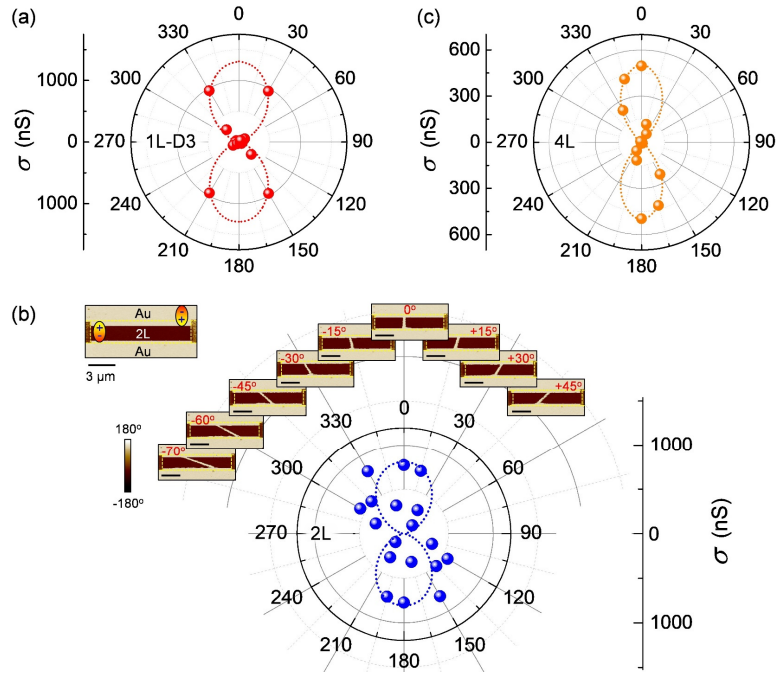

FIG. 3. Polar plots of $\sigma$ of (a) a 1L ReS$_2$ at $V_{bg}$ = 40 V, (b) a 2L ReS$_2$ at $V_{bg}$ = -40 V, and (c) a 4L ReS$_2$ at $V_{bg}$ = 10 V. The dashed lines serve as the guide to the eye. Insets in (b): PFM phase images of the corresponding $P_{down}$ nanowires. The top left image corresponds to the initial uniform $P_{up}$ domain before the nanowire patterning. All scale bars are 3 μm.

Figure 4

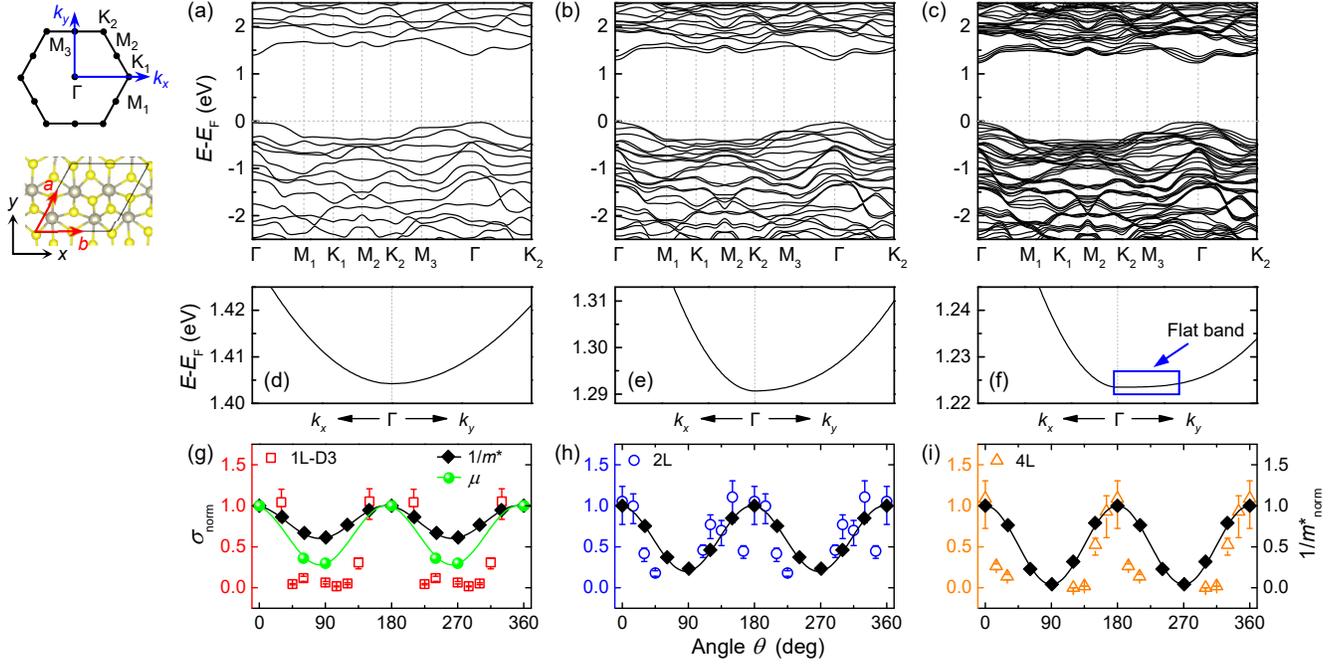

FIG. 4. (a-c) Band structures of (a) 1L, (b) 2L, and (c) 4L ReS$_2$. Insets in (a) show the symmetry points in the first Brillouin zone (top), where $k_x$ and $k_y$ are the same as $x$- and $y$-directions in the real space unit cell (bottom), with (d-f) the corresponding expanded views near the $\Gamma$ point. (g) $\sigma_{norm}$ vs. $\theta$ (open symbols) for the 1L ReS$_2$ in Fig. 3(a) superimposed with the normalized $1/m^*_{norm}$ and $\mu_{norm}$ (solid symbols). (h and i) $\sigma_{norm}$ vs. $\theta$ (open symbols) for (h) the 2L ReS2 in Fig. 3(b) and (i) 4L ReS$_2$ in Fig. 3(c) superimposed with $1/m^*_{norm}$ (solid symbols).

**Giant Transport Anisotropy in ReS$_2$ Revealed via Nanoscale Conducting Path Control (Supplemental Material)**

Dawei Li,[1] Shuo Sun,[1] Zhiyong Xiao,[1] Jingfeng Song,[1] Ding-Fu Shao,[1] Evgeny Y. Tsymbal,[1] Stephen Ducharme,[1] Xia Hong[1*]

[1] Department of Physics and Astronomy & Nebraska Center for Materials and Nanoscience, University of Nebraska-Lincoln, Lincoln, Nebraska 68588-0299, USA

1. Characterization of ReS$_2$ layer thickness and crystalline orientation
2. Characterization of ReS$_2$ transistors and P(VDF-TrFE)/ReS$_2$ samples
3. Ferroelectric field effect in 1L, 2L, and 4L ReS$_2$ FETs
4. Characterization of 1L ReS$_2$ device 1L-D3
5. Directional transport measurements of 2L and 4L ReS$_2$ FETs
6. Theoretical calculations

## 1. Characterization of ReS$_2$ layer thickness and crystalline orientation

We determine the layer thickness and crystalline orientation of ReS$_2$ flakes via Raman spectroscopy. Raman measurements were performed using a micro-Raman system (Renishaw InVia™ Plus) with a 514.5 nm Ar$^+$ laser at a power of ~1.5 mW as excitation light source. For polarized Raman measurements, we collected parallel polarization Raman signal by rotating the sample angle from 0º to 360º in 10º steps.

Figure S1(a) shows the optical image of the 1L ReS$_2$ device 1L-D1 [Fig. 1(a) in the main text]. The layer thickness was identified by combing AFM [Fig. S1(b)] and Raman frequency difference Δ between mode I (at 134.4 cm$^{-1}$) and mode III (at 150.9 cm$^{-1}$) [Fig. S1(c)]. The crystalline orientation of ReS$_2$ was determined by angle-resolved parallel-polarized Raman measurement [Figs. S1(d) and S1(e)], where the *b*-axis is along the channel orientation (*x*-direction).

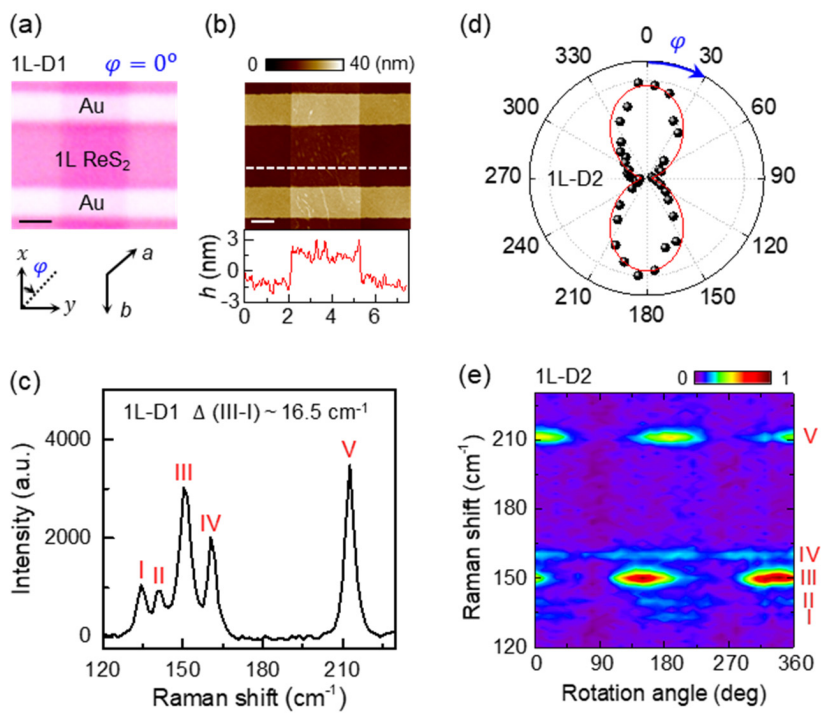

FIG. S1. (a) Optical image of a 1L ReS$_2$ FET, and (b) the corresponding AFM topography image with the height profile along the dashed line. Lower insets in (a) show the laboratory coordinates (left) and crystalline orientation of ReS$_2$ (right). The scale bars are 1 μm. (c) Raman spectrum for the same 1L ReS$_2$ showing mode I at 134.4 cm$^{-1}$ and mode III at 150.9 cm$^{-1}$. (d) Polar plot of Raman intensity for mode V as a function of sample rotation angle $\varphi$ taken on a 1L ReS$_2$, and (e) the corresponding angle resolved Raman spectra with parallel polarized collection.

## 2. Characterization of ReS₂ transistors and P(VDF-TrFE)/ReS₂ samples

Figure S2(a) shows the room-temperature transfer curves $I_d$ vs. $V_{bg}$ for a few-layer (5L) ReS$_2$ device tuned by the SiO$_2$ back-gate. The device exhibits *n*-type switching characteristic behavior before the deposition of P(VDF-TrFE) top layer. We extracted the field effect mobility of the sample from the linear region of the $I_d$-$V_{bg}$ curve using the expression $\mu = \frac{L}{W}\frac{1}{C_{bg}V_d}\frac{dI_d}{dV_{bg}}$, where $L$ ($W$) is the channel length (width), and $C_{bg} = \frac{\varepsilon_0 \varepsilon_r}{d}$ ($\varepsilon_r = 3.9, d = 290\ nm$) is the back-gate capacitance per unit area. The extracted electron mobility is ~13 cm$^2$V$^{-1}$s$^{-1}$, which is similar to previously reported results [1]. After the deposition of a 9 monolayer (ML) P(VDF-TrFE) thin film on top via the Langmuir-Blodgett (LB) technique [2], the transfer curve shifts to the positive $V_{bg}$ direction [Fig. S2(a)], corresponding to a reduced electron doping.

The as-deposited LB film was annealed at 135 °C for 80 minutes to improve its crystallinity [3]. Figures S2(b)-(c) show the piezoresponse force microscopy (PFM) switching hystereses taken on a 9 ML P(VDF-TrFE)/ReS$_2$ sample after thermal annealing. The coercive voltages are about +1.2 V and -1.4 V for the polarization down ($P_{down}$) and up ($P_{up}$) state, respectively, corresponding to a low coercive field of less than 0.09 V/nm.

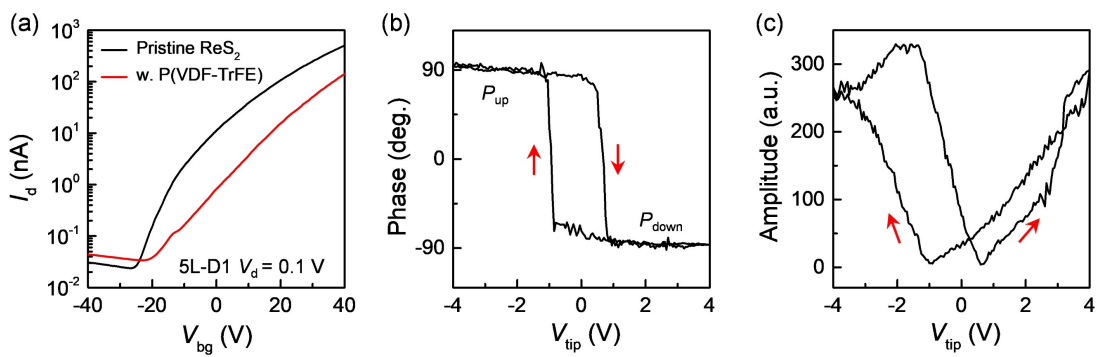

FIG. S2. (a) $I_d$ vs. $V_{bg}$ for a 5L ReS$_2$ FET at $V_d$ = 0.1 V before (black) and after (red) a 9 ML P(VDF-TrFE) thin film deposited on top. (b)-(c) PFM hystereses of (b) phase and (c) amplitude responses taken on a 9 ML LB film on ReS$_2$.

## 3. Ferroelectric field effect in 1L, 2L, and 4L ReS₂ FETs

Figures S3(a)-(c) show the $I_d$-$V_d$ relations at $V_{bg}$ = 0 V taken on 1L, 2L, and 4L ReS$_2$ FETs with the P(VDF-TrFE) top layer uniformly polarized in the $P_{down}$ and $P_{up}$ states, which exhibit a switching ratio of 7.7×10$^4$, 3.2×10$^5$, and 1.3×10$^6$, respectively. Figure S3(d) shows the transfer

curve $I_d$ vs. $V_{bg}$ for the 2L ReS$_2$ FET in Fig. S3(b) in the no poling, $P_{down}$, and $P_{up}$ states of P(VDF-TrFE). In the $P_{down}$ state, the 2L ReS$_2$ channel remains *n*-type and highly conductive over the entire $V_{bg}$–range, similar to that observed in 1L [Fig. 1(f)] and 4L [Fig. 1(g)] ReS$_2$ FETs in the main text. In the $P_{up}$ state, the transfer curve changes to *p*-type characteristic at $V_{bg} < 8$ V. A current switching ratio between the two polarization states reaches $1.9 \times 10^7$ at $V_{bg} = 8$ V, which also corresponds to a change of carrier type.

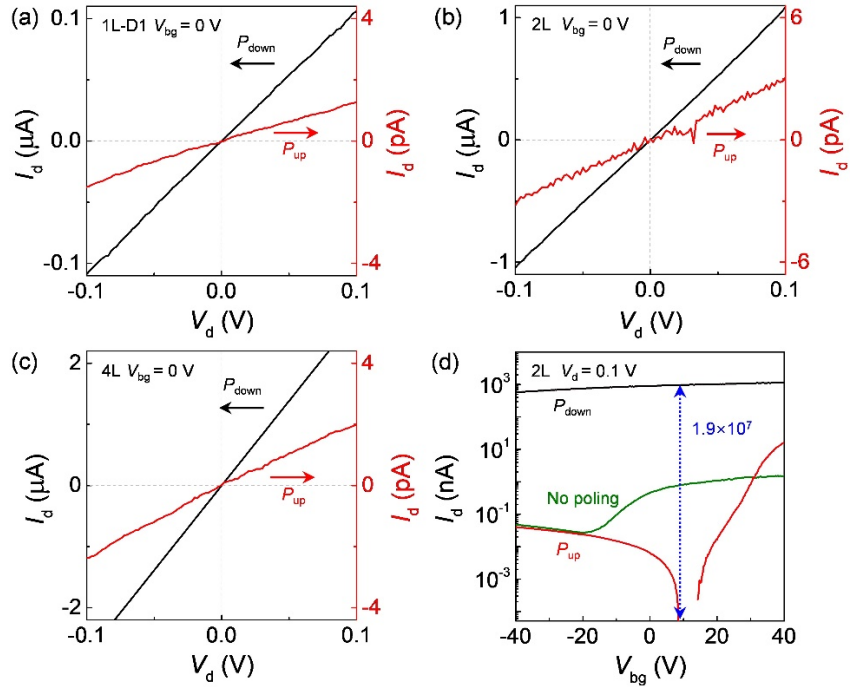

FIG. S3. (a)-(c) $I_d$ vs. $V_d$ at $V_{bg} = 0$ V for (a) 1L, (b) 2L, and (c) 4L ReS$_2$ FETs in the $P_{down}$ and $P_{up}$ states of the P(VDF-TrFE) top-gate. (d) $I_d$ vs. $V_{bg}$ for the 2L ReS$_2$ FET in (b) at $V_d = 0.1$ V in $P_{down}$, $P_{up}$, and no poling states of P(VDF-TrFE).

## 4. Characterization of 1L ReS$_2$ device 1L-D3

Figure S4(a) shows the optical image for the 1L ReS$_2$ device 1L-D3 (Fig. 2 in the main text). The layer thickness was identified using the Raman frequency difference Δ between mode I (at 133.9 cm$^{-1}$) and mode III (at 151.0 cm$^{-1}$) [Fig. S4(b)]. The crystalline orientation of ReS$_2$ was determined by angle-resolved parallel-polarized Raman measurement [Fig. S4(c)], which shows that the *b*-axis (*x*-direction) is perpendicular to the channel orientation (*y*-direction) [Fig. S4(a) insets]. It is also noted that this sample is connected with a thick ReS$_2$ flake, which is cleaved along the crystalline *a*- and *b*-axes, making inner angles of 60° and 120° [Fig. S4(a)].

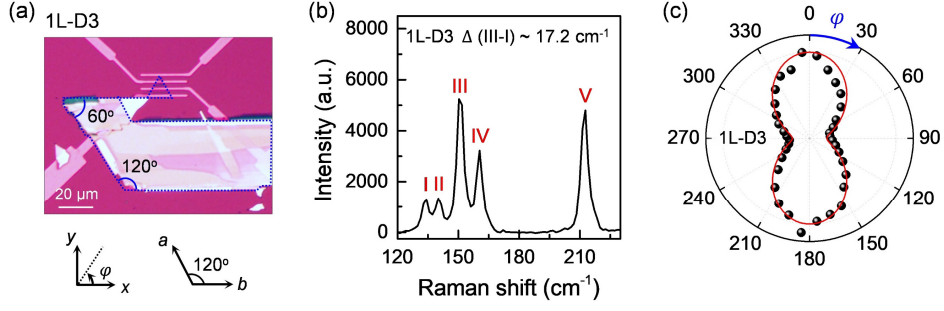

FIG. S4. (a) Optical image for device 1L-D3 shown in Fig. 2 in the main text. The edges of the monolayer ReS$_2$ flake are outlined. The lower insets show the laboratory coordinate system (left) and crystalline axes of ReS$_2$ (right). (b) Raman spectrum for the 1L ReS$_2$ in (a) showing mode I at 133.9 cm$^{-1}$ and mode III at 151.0 cm$^{-1}$. (c) Polar plot of the Raman intensity for mode V as a function of sample rotation angle $\varphi$.

## 5. Directional transport measurements of 2L and 4L ReS$_2$ FETs

Figure S5(a) shows the sheet conductance $\sigma$ as a function of $V_{bg}$ for the directional nanowires created in the 2L ReS$_2$ FET. The corresponding PFM images are shown in Fig. 3(b) in the main text. The device exhibits *n*-type transfer characteristic along all directions over the entire $V_{bg}$-range. To ensure the Fermi level is close to the conduction band edge, we extracted the angle-resolved conductivity at $V_{bg}$ = -40 V [Fig. 3(b) in the main text], which corresponds to the lowest electron doping level.

Figure S5(b) shows the sheet conductance $\sigma$ as a function of $V_{bg}$ for the directional nanowires created in the 4L ReS$_2$ FET shown in Fig. 3(c) in the main text. The conductivity anisotropy is strongly enhanced in this sample. Here, we observed a giant anisotropy ratio $\sigma_{max}(\theta = 0°)/\sigma_{min}(\theta = 120°)$ of 6.7×10$^3$ at $V_{bg}$ = -14 V, which also corresponds to a change of carrier type from *n*- to *p*-type. In Fig. 3(c), we extracted the angle-resolved conductivity at $V_{bg}$ = 10 V, which corresponds to *n*-type conduction for all nanowires with the lowest electron doping level.

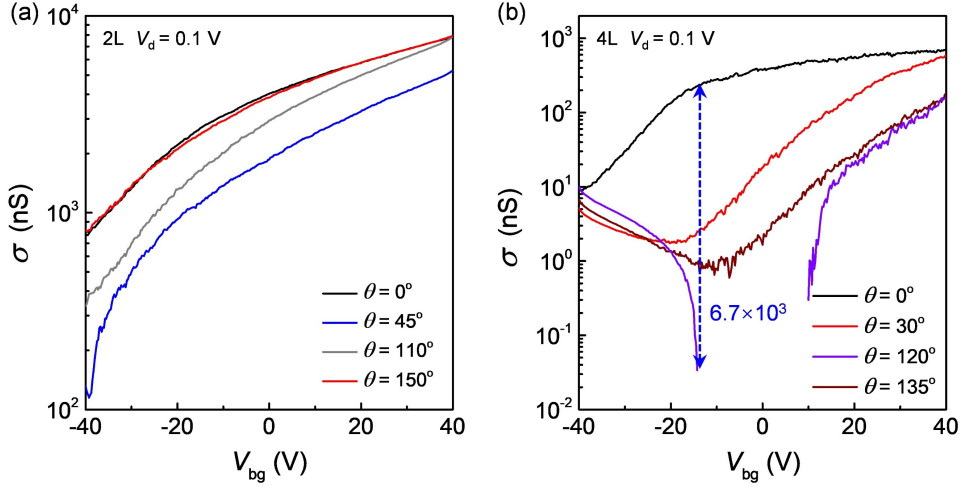

FIG. S5. Selected $\sigma$ vs. $V_{bg}$ curves for (a) 2L and (b) 4L ReS$_2$ FETs at $V_d = 0.1$ V measured with the conducting nanowires along different orientations $\theta$ with respect to the $b$-axis.

## 6. Theoretical calculations

First-principles calculations were performed with the projector augmented-wave (PAW) method [4] implemented in the VASP code [5]. The exchange and correlation effects were treated within the generalized gradient approximation (GGA) [6]. We used the plane-wave cut-off energy of 550 eV and a $16 \times 16 \times 1$ $k$-point mesh in the irreducible Brillouin zone. A vacuum layer over 25 Å was used in the unit cell to separate the top and bottom surface of ReS$_2$. The van der Waals corrections as parameterized in the semiempirical DFT-D3 method [7] were included in the calculation of 4L ReS$_2$. The deformation potential $D_i$ was calculated using strain varying from $-1\%$ to $1\%$ with the step of $0.5\%$. Figure S6 shows the calculated $1/m^*$ vs. $\theta$ relation for 1L [Fig. S6(a)], 2L [Fig. S6(b)] and 4L [Fig. S6(c)] ReS$_2$.

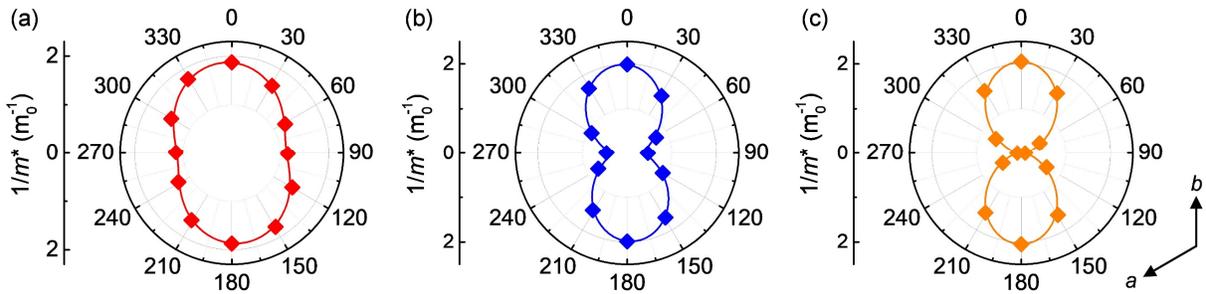

FIG. S6. Polar plots of $1/m^*$ vs. $\theta$ for (a) 1L, (b) 2L, and (c) 4L ReS$_2$ in the unit of free electron mass $m_0$.